\documentclass[%
reprint,
superscriptaddress,
 amsmath,amssymb,
 aps,
prl,
]{revtex4-1}

\usepackage{graphicx}% Include figure files
\usepackage{dcolumn}% Align table columns on decimal point
\usepackage{bm}% bold math
\begin{document}

% Use the \preprint command to place your local institutional report
% number in the upper righthand corner of the title page in preprint mode.
% Multiple \preprint commands are allowed.
% Use the 'preprintnumbers' class option to override journal defaults
% to display numbers if necessary
%\preprint{}

%Title of paper
\title{Suppression of $f$-Electron Itinerancy in CeRu$_2$Si$_2$ by a Strong Magnetic Field}

% repeat the \author .. \affiliation  etc. as needed
% \email, \thanks, \homepage, \altaffiliation all apply to the current
% author. Explanatory text should go in the []'s, actual e-mail
% address or url should go in the {}'s for \email and \homepage.
% Please use the appropriate macro foreach each type of information

% \affiliation command applies to all authors since the last
% \affiliation command. The \affiliation command should follow the
% other information
% \affiliation can be followed by \email, \homepage, \thanks as well.
\author{Y.~H.~Matsuda}
\email{ymatsuda@issp.u-tokyo.ac.jp}
\affiliation{Institute for Solid State Physics, University of Tokyo, Kashiwa, Chiba 277-8581, Japan}
%\homepage[]{Your web page}
%\thanks{}
%\altaffiliation{}

\author{T.~Nakamura}
\affiliation{Institute for Solid State Physics, University of Tokyo, Kashiwa, Chiba 277-8581, Japan}

\author{J.~L.~Her}
\altaffiliation[Present address, ]{Center for General Education, Chang Gung University, Taoyuan County 333, Taiwan (R.O.C.)}
\affiliation{Institute for Solid State Physics, University of Tokyo, Kashiwa, Chiba 277-8581, Japan}

\author{S.~Michimura}
\author{T.~Inami}
% \homepage{http://www.Second.institution.edu/~Charlie.Author}
\affiliation{Condensed Matter Science Division, Japan Atomic Energy Agency, Sayo, Hyogo 679-5148, Japan}%

\author{K.~Kindo}
% \homepage{http://www.Second.institution.edu/~Charlie.Author}
\affiliation{Institute for Solid State Physics, University of Tokyo, Kashiwa, Chiba 277-8581, Japan}

\author{T.~Ebihara}
% \homepage{http://www.Second.institution.edu/~Charlie.Author}
\affiliation{Department of Physics, Faculty of Science, Shizuoka University, Shizuoka 422-8529, Japan}

%\author{H.~Amitsuka}
% \homepage{http://www.Second.institution.edu/~Charlie.Author}
%\affiliation{Graduate School of Science, Hokkaido University, Sapporo 060-0810, Japan}

%Collaboration name if desired (requires use of superscriptaddress
%option in \documentclass). \noaffiliation is required (may also be
%used with the \author command).
%\collaboration can be followed by \email, \homepage, \thanks as well.
%\collaboration{}
%\noaffiliation

\date{\today}

\begin{abstract}
The valence state of Ce in a canonical heavy fermion compound CeRu$_2$Si$_2$ has been investigated by synchrotron X-ray absorption spectroscopy at 1.8~K in high magnetic fields of up to 40~T. The valence was slightly larger than for the pure trivalent state (Ce$^{3+}$: $f^1$), as expected 
in heavy fermion compounds, and it decreased toward the trivalent state as the magnetic field was increased. 
%when the field is higher than the metamagnetic transition field ($H_m \sim$ 8~T).
%Since the itinerant character of 4$f$ electrons manifests itself in the deviation of the occupation number ($n_f$)
%of the 4$f$-orbital from unity,  i.e., $1-n_f$, 
The field-induced valence reduction indicates that the itinerant character of the 4$f$ electrons in 
CeRu$_2$Si$_2$ was suppressed by a strong magnetic field. 
The suppression was gradual and showed characteristic magnetic field dependence, which reflects the metamagnetism around $H_m \sim$ 8~T. 
The itinerant character persisted, even at 40~T ( $\sim5H_m$), suggesting that the Kondo bound state is continuously broken by magnetic fields and that it %should 
may completely collapse at fields exceeding 200~T. 

\end{abstract}

% insert suggested PACS numbers in braces on next line
% insert suggested keywords - APS authors don't need to do this
%\keywords{}

%\maketitle must follow title, authors, abstract, \pacs, and \keywords
\maketitle

% body of paper here - Use proper section commands
% References should be done using the \cite, \ref, and \label commands
%\section{Introduction}
%##
The correlation between localized and itinerant electrons is one of the most intriguing subjects in condensed matter physics. 
Heavy electrons emerge in rare-earth and actinide compounds and give rise to exotic phenomena such as non-Fermi liquid behavior 
\cite{Stewart01, Lohneysen07, Gegenwart08} 
and unconventional superconductivity \cite{Grosche98, Monthoux07, Gegenwart08} because of their strong correlation.
%##
%Recently, the quantum criticality of the valence transition (VT) of the rare-earth elements has been attracting much interest 
%in addition to the magnetic quantum criticality. \cite{} 
%Superconductivity is discovered in CeCu$_2$Si$_2$ at high pressures \cite{}  where the valence fluctuation of Ce 
%is enhanced. \cite{} 
%Moreover, it is theoretically suggested that the quantum critical point (QCP) of the VT is induced by applying high magnetic fields and 
%the metamagnetism can be observed at near the QCP. \cite{}
%
%The quantum criticality is the key to understand theses unusual phenomena.
%The quantum critical point (QCP) due to the competition between the localized magnetic ordered state through the 
%Ruderman-Kittel-Kasuya-Yoshida (RKKY) interaction and the itinerant nonmagnetic state owing to the Kondo effect 
%has been investigated enthusiastically in the heavy fermion (HF) systems.
The magnetic quantum critical point (QCP) in the Doniac model \cite{Doniach77, Lohneysen07} 
is one of the most important concepts for understanding the low-temperature physics of heavy fermion (HF) systems. However, another type of QCP, which is caused by 
the valence transition (VTQCP),\cite{Miyake07, Watanabe08} has been proposed as the origin of the unusual superconductivity, 
such as that observed in CeCu$_2$Si$_2$ \cite{Rueff11} and also as the origin of the metamagnetism inYb-based HF systems \cite{Watanabe08,Watanabe10}.
%another type quantum criticality was recently proposed as the origin of the unusual phenomena in heavy fermions. 
%It is due to the charge degree of freedom instead of the spin degree of freedom, i.e., QCP of the valence transition (VTQCP).\cite{Miyake07, Watanabe08} 
%The typical case is the superconductivity in CeCu$_2$Si$_2$ where the fluctuation in the Ce valence is enhanced by 
%pressures. \cite{Rueff11} 
%Moreover, the magnetic field induced QCP found in Yb-based heavy fermion compounds can also be understood in terms of the valence 
%fluctuation phenomena.\cite{Watanabe10}f
%Investigation of the valence state in heavy fermion systems has attracted considerable attention. \cite{Okawa10, Matsumoto11}

Most of the HF systems contain Ce or Yb. 
The occupation number of the 4$f$ electron in the orbital ($n_f$) is close to 1 for Ce, and
in Yb, the occupation number of the 4$f$ hole ($n_h$) is $\sim1$. 
The 4$f$ electrons acquire an itinerant character because of the strong hybridization with the conduction electrons.
The energy level of the correlated 4$f$ electrons is near the Fermi energy, because the 
significant electron-electron many-body interaction causes the electrons to become itinerant HFs. %\cite{} 
The degree of the itinerancy is determined by the strength of the electron interaction. The itinerant character manifests itself 
in the deviation of the occupation number of the $f$ electron or hole from unity:  $1-n_f$ or $1-n_h$.
According to the G\"{u}tzwiller approximation, the deviation $1-n_f$ can be expressed by 
$1-n_f = \frac{kT_F^*}{\Gamma}$, where $kT_F^*$ and $\Gamma$ correspond to the kinetic energy of the $f$ electrons with the 
electron interaction and that without the interaction, respectively.\cite{heavy}
In HF systems, $1-n_f$ must be very small but finite, indicating that the electrons are highly localized and only slightly itinerant.

The direct observation of the valence of Ce and Yb in heavy fermion compounds in high magnetic fields is particularly desirable because the magnetic field dependence of $n_f$ and the change in the itinerant behavior of the electrons can be directly measured. 
However, it is not easy to determine the precise magnetic field variation of $1-n_f$ or $1-n_h$ experimentally, because they generally have values smaller than 0.1 in HF systems; the field dependence is expected to be of the order of 0.01 or smaller.
 Therefore, the change in the behavior of the heavy electrons caused by a magnetic field has rarely been studied in terms of the valence state.
%except for several materials that show significant valence fluctuating phenomena such as YbInCu$_4$. \cite{} 

CeRu$_2$Si$_2$ is a canonical HF system and has been attracting considerable attention for its metamagnetism 
around $H_m \sim$ 8~T.\cite{Haen87} 
It was proposed that the transformation of the 4$f$ electrons from the itinerant to the localized state induced by a magnetic field (so-called the Kondo breakdown) 
could be observed based on the change from a large to small Fermi surface. \cite{Aoki93}
However,  it was later found that the change in the Fermi surface is not necessarily concrete evidence of the transformation of the $f$ electrons \cite{Watanabe00, MiyakeIkeda06, Daou06}; the possible change in the behavior of the $f$ electrons is still controversial. \cite{Sugi08} 

We have performed synchrotron X-ray absorption spectroscopy for CeRu$_2$Si$_2$ in high magnetic fields of up to 40~T at 1.8~K.
The magnetic field caused significant changes in the X-ray absorption spectrum near the Ce-$L_3$ edge, suggesting a change in the Ce valence toward the pure trivalent state.
%We have successfully determined the magnetic field dependence of the Ce valence at 1.8~K and found that the itinerancy of the 4$f$ electrons is 
%suppressed by magnetic fields.  
The suppression of the $f$-electron itinerancy and the magnetic field dependence %were also examined. 
are discussed.
%
%これまでのメタマグの説明と整合するか？　Kondo breakdown, spin split Fermi surface . . . 
%遍歴性の減少は、ミクロにどう捉えられるか？　混成Vの減少？　
%最後の議論の部分では量子臨界点についても言及すべき、電気伝導との関連は？
%
%The Ce valence determined by the XAS is slightly higher than the pure trivalent state (Ce$^{3+}$: $f^1$) at 1.8~K and zero magnetic field.
%It is found that the valence decreases with magnetic fields when the magnetic field exceeds the metamagnetic transition field ($H_m \sim$ 8~T). 
%Moreover, the valence is fluctuating even at 40~T, suggesting the 4$f$ electron itinerancy remains at even this high magnetic fields.
%

%\section{Experiments}
The X-ray absorption experiment was carried out at beamline BL22XU of SPring-8. 
Pulsed high magnetic fields of up to 40 T were generated by a miniature pulsed magnet.\cite{Matsuda07}
The duration of the magnetic field was ~1 ms and the repetition rate was 4 pulses per hour for a 40~T measurement. 
The magnet and sample were cooled to 1.8 K by using an ILL-type He gas flow cryostat.
The X-ray absorption spectrum was taken by a direct transmission method.
The details of the experimental techniques have previously been described. \cite{Matsuda07}
A single crystal of CeRu$_2$Si$_2$ was grown by the Czochralski pulling method.
The crystals were powdered and mixed with epoxy resin so that the effective thickness was appropriate for an X-ray absorption 
intensity of $\mu t \sim1$, where $\mu$ and $t$ are the absorption coefficient and the thickness of the sample, respectively.
The $c$-axes of the powder crystals were aligned by means of a steady magnetic field of 14~T when the epoxy resin was solidified.
The diameter of the powder was $\mu$m. 
Diluting fine powders in epoxy resin avoided Joule heating of the sample by a pulsed magnetic field.
The high magnetic field magnetization data was obtained by the induction method using a conventional pulsed magnet. 

\begin{figure}[b]
\includegraphics[width=7cm]{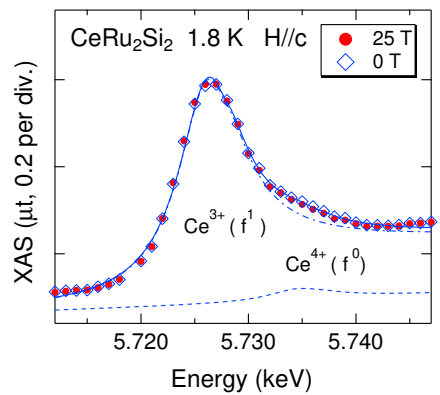}% Here is how to import EPS art
\caption{\label{fig:XAS} X-ray absorption spectra near the Ce $L_3$ edge at 0 and 25~T.
The solid, dotted, and dot-dashed curves are the results of the spectrum fitting to the zero magnetic field spectrum.}
\end{figure}

%\section{Results of Experiments}
The X-ray absorption spectra (XAS) at 1.8~K in a zero magnetic field and at 25~T are shown in Fig.~\ref{fig:XAS}.
The magnetic field was applied in parallel to the $c$-axis of the sample. 
Although there is a clear magnetic field effect, which is discussed later, the spectrum at 25~T was very similar to that at 0~T. 
The absorption peak near 5.727~keV was the white line at the Ce $L_3$ absorption edge in the trivalent state (Ce$^{3+}$: $f^1$).
A small absorption band was observed in the spectrum around 5.735~keV, which can be attributed to the tetravalent state (Ce$^{4+}$: $f^0$) \cite{Rueff06}. 
The spectrum shape fitting was performed using the standard Lorentz and arctangent functions with a linear background.
The solid curve, dotted, and dot-dashed curves are the results of the spectrum fitting to the zero magnetic field curve.
The dot-dashed and dotted curves represent the $f^1$ and $f^0$ states, respectively, and the solid curve represents the whole shape of the spectrum. 
The valence $v$ was directly determined by the relative intensity of the absorption bands, $v$ = 3+$I$($f^0$)/($I$($f^0$)+$I$($f^1$)), where 
$I$($f^0$) denotes the absorption intensity of the $f^0$ absorption band and $I$($f^1$) denotes that of the $f^1$ absorption band.  
%%%%
%%%%%YHM
The observed $f^0$ and $f^1$ contributions are 0.053 and 0.947, respectively, 
and the Ce valence was calculated as $v=3.053 \pm 0.02$ ($1-n_f = 0.053\pm 0.02$) at zero magnetic fields.

%YHM
The $f^0$ component of CeRu$_2$Si$_2$ at 20 K was reported to be 0.06 by the photoemission spectroscopy and 
the soft-X-ray absorption spectra with the analysis using the impurity Anderson model. \cite{yano08, willers12}
Hence, the estimation of the $f^0$ component in our XAS study at 1.8~K seems to be in good agreement with the 
previous reported value.
Since we did not take $f^2$ contribution into account in this work, we may overestimate $n_f$. 
Actually $n_f$ was deduced to be 0.013 at 20~K %YHM
%considering $f^2$ contribution (0.05) in the previous works. \cite{yano08, willers12}
if we assume that there is the $f^2$ contribution of 0.05. \cite{yano08, willers12}  %FPRB
However,%YHM
in terms of the Kondo breakdown scenario, the $f^0 \rightarrow f^1$ transformation is the dominant magnetic field effect.
Moreover, we can safely assume that the $f^2$ state is much less sensitive to magnetic field than the $f^1 \leftrightarrow f^0$ 
transformation similarly to the pressure variation of Ce valence in CeCu$_2$Si$_2$. \cite{Rueff11}
% the $f^2$ contribution and the magnetic field dependence were not observed in the present study within the 
%experimental error, and we can safely assume that the $f^2$ state is much less sensitive to magnetic field than the $f^1 \leftrightarrow f^0$ 
%transformation similarly to the pressure variation of Ce valence in CeCu$_2$Si$_2$. \cite{Rueff11}
%In the present study we simply assume that $n_f=1$ when the f-electron becomes completely localized. 
%If we consider the $f^2$ contribution $n_f$ can exceed the unity as in the case of CeRu$_2$Ge$_2$.\cite{yano08}
%The correction by an almost constant $f^2$ contribution is considered to be a normalization factor.
%Since the $f^0 \rightarrow f^1$ transformation is the dominant magnetic field effect,  the inclusion of the $f^2$ contribution (about 0.05  \cite{yano08}) 
%cannot be important for evaluation of the field-induced valence change unless $f^2$ contribution shows an unreasonably large field dependence.
%Since $f^2$ contribution is as large as 0.04 and there is no physical reason for 
Therefore, since $1- n_f$ obtained in this work corresponds to the $f^0$ component and should directly reflects the degree of the itinerancy, 
a small correction by the $f^2$ component is not important for evaluation of the magnetic field effect on the valence state in the present study.

\begin{figure}[b]
\includegraphics[width=6 cm]{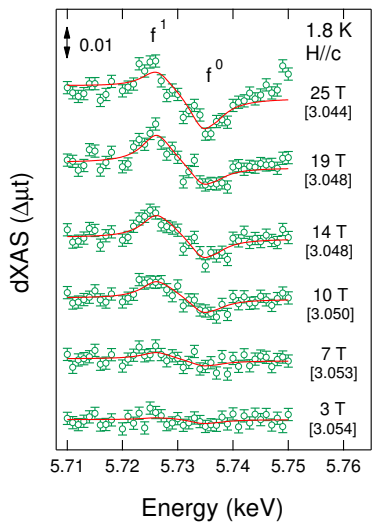}% Here is how to import EPS art
\caption{\label{fig:dXAS} Difference spectra obtained by subtracting the XAS at zero magnetic fields from that at finite magnetic fields.
The open circles are the experimental results and the solid curves were obtained from the XAS fitting. The values in the parentheses are the valences deduced from the fitting analysis. The relative error from the zero field valence ($v = 3.053$) was about $\pm 0.003$.
%Two vertical dashed lines denote the energy positions where the 10 times repetition measurement was conducted; the results are shown in Fig.~\ref{fig:B_Dep}.
}
\end{figure}

The change in the spectrum induced by the magnetic field was very small; therefore the spectrum at zero magnetic field was subtracted from the spectrum at finite magnetic fields in order to see the field evolution of the spectrum clearly. 
The difference XAS (dXAS) at different magnetic fields are shown in Fig.~\ref{fig:dXAS}.
The open circles are the experimental results and the solid curves are derived from the fitting curves.
The positive and negative peak structures appeared around 5.725 and 5.735~keV, respectively; these energy positions correspond to the absorption bands from the $f^1$ and $f^0$ states shown in Fig.~\ref{fig:XAS}. 
This characteristic feature evolved as the magnetic fields increased, suggesting the $f^1$ state became more prominent, whereas the $f^0$ state 
diminished. 
The intensities of the $f^1$ and $f^0$ band were used as the fitting parameters and the fitting results accurately reproduced the characteristic features 
of the dXAS (Fig.~\ref{fig:dXAS}). 
Therefore, the valence decreases toward the trivalent state as the magnetic field increases.
%Note that this is a distinct evidence of the change in the valence state by magnetic fields. 
When the same experiment was performed with the magnetic field perpendicular to the $c$-axis ($H \perp c$), 
the dXAS was almost flat with no significant features even at 25~T, suggesting that there was no field dependence of the valence state when $H \perp c$.
%
%YHM
Since the magnetization for $H \perp c$ is more than 10 times smaller than that for $H \parallel c$, it is expected that the magnetic energy gain for 
$H \perp c$ is not sufficient to induce a suppression of the $f$-electron itinerancy. 
The Ce valence was evaluated at different magnetic fields through the fitting analysis, and decreased as the magnetic field increased.  
The valences are shown in parentheses in Fig.~\ref{fig:dXAS}. The experimental error in the relative change of the valence from the zero field value was about $\pm 0.003$, which is too large to allow for detailed discussion of the magnetic field dependence of the valence.

\begin{figure}[b]
\includegraphics[width=6.5 cm]{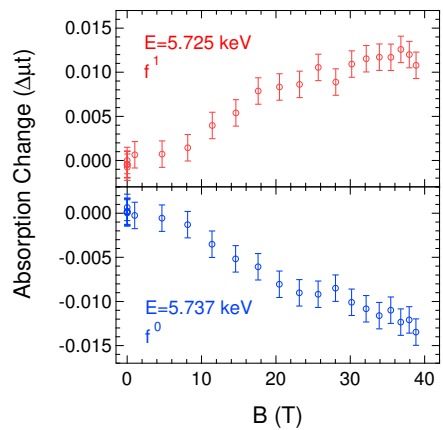}% Here is how to import EPS art
\caption{\label{fig:B_Dep} Change in the absorption intensity at 5.725 and 5.737~keV induced by magnetic fields. }
\end{figure}

To observe the magnetic field variation of the valence more precisely, we focused on two particular energy positions, 5.725 and 5.737~keV. The measurements for the magnetic field dependence of the absorption intensity were repeated 10 times for each energy and the average was taken (Fig.~\ref{fig:B_Dep}). 
The change in the absorption intensities at the two energy positions should follow the change in the components of the $f^1$ and $f^0$ states. 
It is found that the change in the absorption intensity at low magnetic fields was small and the change rate increased under magnetic fields around 8~T. 

\begin{figure}[b]
\includegraphics[width=6.5 cm]{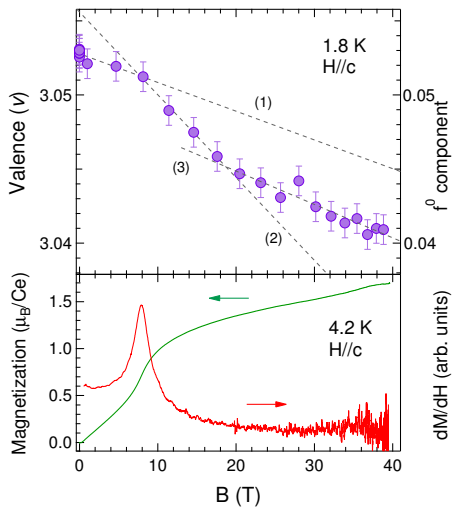}% Here is how to import EPS art
\caption{\label{fig:valence} (a) Magnetic field dependence of the Ce valence and the corresponding $f^0$ component at 1.8~K. 
The dashed lines (1), (2) and (3) show the slope of the valence 
change in low, medium and high magnetic field ranges, respectively. (b) Magnetic field dependence of the magnetization ($M$) and the field derivative 
($dM/dH$) at 4.2~K.}
\end{figure}

Assuming that only the intensity changed and that the shape and the energy shift of the $f^1$ and $f^0$ absorption bands were not altered, 
the magnetic field dependence of the valence was determined from the results shown in Fig.~\ref{fig:B_Dep}. 
In Fig.~\ref{fig:valence} (a), the valence is plotted as a function of magnetic field, and the magnetization ($M$) and its magnetic field derivative ($dM/dH$) at 4.2 K are shown as a function of the magnetic field in Fig.~\ref{fig:valence} (b) 
for comparison. 
In the $M$ and $dM/dH$ curves, the metamagnetic transition is clearly visible at 8~T ($H_m$). 
The valences were in good agreement with those obtained by XAS shown in Fig.~\ref{fig:dXAS} within the experimental error.
Because of the better signal to noise ratio of the results shown in Fig.~\ref{fig:valence}, 
the unusual magnetic field dependence of the valence was visible. The valence decreased slightly in low magnetic fields and the rate of the decrease was larger when the magnetic field was higher than about 8~T; this magnetic field was defined as $H_v$. 
The two dashed lines labeled (1) and (2) show the slope of the valence change in low and high magnetic fields around $H_v$.
It was clear that $H_v$ corresponded to the metamagnetic transition field, $H_m$. 
Because $H_m$ shows almost no temperature dependence, \cite{Haen87} the temperature difference between 1.8 and 4.2~K 
%should not be important. 
is not important as we compare $H_v$ and $H_m$. %FPRB
The valence gradually decreased and continued to change even at magnetic fields higher than $H_m$.
The valence decreased even at 40~T, although the metamagnetic transition was probably complete above 20~T. 
%As we above mentioned no magnetic field effect on the valence state was observed when $H \perp c$.
%Therefore, it is experimentally obvious that the valence begins to decrease significantly at higher magnetic fields beyond the metamagnetism.
%
%

%\section{Discussion}
The significance of the decrease in the Ce valence when the magnetic field was applied parallel to the $c$-axis of the crystal was examined. 
It is generally accepted that the HF state is broken if a strong enough magnetic field is applied. 
This is because the heavy fermion state is reached through the Kondo singlet bound state, and the singlet state can be broken if the Zeeman energy 
exceeds $k_BT_K$, where the $k_B$ is the Boltzmann constant and $T_K$ is the Kondo temperature. 
For CeRu$_2$Si$_2$, shown in Fig.~\ref{fig:valence}, the corresponding 
Zeeman energy is about 14~K which is comparable to the $T_K \sim$ 24~K, \cite{Haen87} %FPRB 
because the magnetization is roughly 1~$\mu_B$ at 10~T. 
Hence, we propose that the Kondo bound state begins breaking at the metamagnetic transition. 
This qualitative explanation is the same as that given for the Fermi surface shrinkage at the metamagnetic transition. \cite{Aoki93} 
However, if the $f$ electron becomes completely localized because the Kondo singlet state is broken, the Ce valence state should become 
purely trivalent and independent of magnetic fields. 
According to our x-ray absorption results, %YHM
%the valence was still slightly larger than 3 
the $f^0$ component was still finite value and continued to decrease with the magnetic field, even at 40~T.
Therefore, it is found that the itinerancy of the electrons is gradually suppressed by the magnetic field 
and the suppression becomes prominent at the metamagnetic transition field $H_m$.
%FPRB
%Therefore, the heavy fermion state was suppressed by the magnetic field and the suppression was increased by the metamagnetism 
%at $H_m$, although the heavy state was not completely broken, even at magnetic fields as strong as $5H_m$.
The %reduction
suppression rate %was 
becomes small again around 19~T where the lines (2) and (3) cross, and the metamagnetic transition also appears to finish 
according to the $dM/dH$ curve.
The suppression continues even at magnetic fields as strong as $5H_m$.%FPRB

%FPRB It is that the 
The valence fluctuation phenomena around $H_m$ in CeRu$_2$Si$_2$ may be related to the QCP of the valence transition (VTQCP), as has been suggested in other heavy fermion compounds.\cite{Watanabe09, Watanabe10}
The Ce valence of CeRu$_2$Si$_2$ was thought to be nearly trivalent and stable in magnetic fields.
However, we have discovered that the valence depends on the applied magnetic field and changes by about 0.005 
around the metamagnetic transition (8-18~T). 
The valence change was several times smaller than that of YbAgCu$_4$ \cite {Matsuda12} a heavy fermion material that exhibits metamagnetism, 
which is probably caused by the VTQCP.
It is not clear at the present how far away CeRu$_2$Si$_2$ is located from the VTQCP; \cite{Watanabe09, Watanabe10} 
and the mechanism of the metamagnetic transition in CeRu$_2$Si$_2$ should be re-examined.

Another interesting finding is that the valence change can quantitatively explain the large magnetovolume effect in CeRu$_2$Si$_2$ \cite{Paulsen90}.
When 5.221 and 4.661 $\textrm{\AA}$ were used for the ionic radii ($r$) of Ce$^{3+}$ and Ce$^{4+}$, respectively, \cite{Grier81} the relative change 
in the radius $\frac{\Delta r}{r}$ was $6.47 \times 10^{-4}$ at 12~T using the valence shown in Fig.~\ref{fig:valence}, 
which gave the relative volume change $\Delta V /V \sim1.94 \times 10^{-3}$. % without consideration of the anisotropy. 
This was consistent with the $\Delta V /V \sim1.8 \times 10^{-3}$ at around 12~T that was obtained from magnetostriction experiments. \cite{Paulsen90}
%YHM
This is the first clear evidence that the magnetovolume effect in CeRu$_2$Si$_2$ is due to the field-induced valence change.%%FPRB
%This find is very impressive because the valence instability has never been addressed as the origin of the magnetovolume effect in CeRu$_2$Si$_2$.

The field variation of the valence observed in this study corresponds to the loss of itinerancy in terms of the G\"{u}tzwiller approximation; 
magnetic fields reduce the value of $1-n_f$.
The relative reduction of the itinerancy was possibly evaluated as $ \frac{\Delta n_f }{1-n_f (0~T)} \sim\frac{0.005}{0.053} \sim0.09$ around 
the metamagnetic transition (8-18~T); $\Delta n_f $ corresponds to the change in the valence in a magnetic field and $n_f = 4 - v$.
If a complete field-induced valence transition takes place, $\frac{\Delta n_f }{1-n_f (0~T)} $ should be close to 1.0.
It is actually about 0.7 in the Yb-based heavy fermion compound, YbAgCu$_4$.\cite {Matsuda12} 
The value of $ \frac{\Delta n_f }{1-n_f (0~T)} \sim0.09$ in CeRu$_2$Si$_2$ suggests that the itinerancy is suppressed by only 9\%.
This is considerably smaller than the value expected for the complete valence transition. 
At 40~T, the value was $ \frac{\Delta n_f }{1-n_f (0~T)} \sim\frac{0.012}{0.053} \sim0.23$, suggesting that the $f$ electrons were still 
far from being completely localized. 
%YHM
This evaluation is not changed if we consider the $f^2$ contribution that we did not take into account in our analysis. 
It is because the localization of the $f$-electron can be evaluated by the reduction of the $f^0$ component.

%Assuming $T_K$ is scaled with $1-n_f$, $T_K$ at 40~T is 18.5~K if we use $T_K$= 24~K \cite{Haen87} at 0~T.
%
At a very high magnetic field, $1-n_f = 0$ and the $f$ electrons should become completely localized. % where $T_K$ should be 0 K.
If the magnetic field, $H_L$, where $1-n_f = 0$ is simply estimated using an extrapolation of line (3) in Fig.~\ref{fig:valence} (a) to the position $v$ = 3.00, 
then $H_L \sim$ 220~T.

%\section{Conclusion}
In conclusion, we have found that the Ce valence fluctuated at 1.8~K in CeRu$_2$Si$_2$ and that it decreased 
toward the pure trivalent state as the magnetic field was increased. 
%The heavy electron state is suppressed by a magnetic field and the suppression rate becomes large at the metamagnetic transition.
The valence reduction, which corresponds to the metamagnetism at $H_m \sim$ 8~T, was around 0.005, and was too small for the metamagnetism to be considered as the complete localization of the electrons. 
The valence changes gradually and thus cannot be a first order transition as has previously been suggested \cite{Sakakibara95, Daou06}. 
The characteristic magnetic field dependence of $n_f$, or the Ce valence, can be understood by the gradual break down of the Kondo bound state.
However, the relationship between the valence change and the metamagnetism is still not clear. 
CeRu$_2$Si$_2$ may be located close to the VTQCP, and it can be drawn closer by applying magnetic fields.
Further theoretical studies are necessary to solve this problem.
%We obtained $1-n_f \sim 0.053$ at zero magnetic field and $\sim 0.041$ at 40~T; the itinerancy weakens by 23\% at 5$H_m$. 
%Complete localization of $f$ electrons in CeRu$_2$Si$_2$ is likely to be realized at very high magnetic fields over 200~T. 

%\section{Aknowledgements}
Y.~H.~M thanks Prof. S.~Watanabe for useful discussions.
This work is partly supported by a Grant-in-Aid for Scientific Research B (22340091) and for Scientific Research A (22244047) provided by the Ministry of Education, Culture, Sports, Science and Technology (MEXT), Japan.

% Create the reference section using BibTeX:
%\bibliography{ce_xas_ref.bib}

\begin{thebibliography}{24}%
\makeatletter
\providecommand \@ifxundefined [1]{%
 \@ifx{#1\undefined}
}%
\providecommand \@ifnum [1]{%
 \ifnum #1\expandafter \@firstoftwo
 \else \expandafter \@secondoftwo
 \fi
}%
\providecommand \@ifx [1]{%
 \ifx #1\expandafter \@firstoftwo
 \else \expandafter \@secondoftwo
 \fi
}%
\providecommand \natexlab [1]{#1}%
\providecommand \enquote  [1]{``#1''}%
\providecommand \bibnamefont  [1]{#1}%
\providecommand \bibfnamefont [1]{#1}%
\providecommand \citenamefont [1]{#1}%
\providecommand \href@noop [0]{\@secondoftwo}%
\providecommand \href [0]{\begingroup \@sanitize@url \@href}%
\providecommand \@href[1]{\@@startlink{#1}\@@href}%
\providecommand \@@href[1]{\endgroup#1\@@endlink}%
\providecommand \@sanitize@url [0]{\catcode `\\12\catcode `\$12\catcode
  `\&12\catcode `\#12\catcode `\^12\catcode `\_12\catcode `\%12\relax}%
\providecommand \@@startlink[1]{}%
\providecommand \@@endlink[0]{}%
\providecommand \url  [0]{\begingroup\@sanitize@url \@url }%
\providecommand \@url [1]{\endgroup\@href {#1}{\urlprefix }}%
\providecommand \urlprefix  [0]{URL }%
\providecommand \Eprint [0]{\href }%
\providecommand \doibase [0]{http://dx.doi.org/}%
\providecommand \selectlanguage [0]{\@gobble}%
\providecommand \bibinfo  [0]{\@secondoftwo}%
\providecommand \bibfield  [0]{\@secondoftwo}%
\providecommand \translation [1]{[#1]}%
\providecommand \BibitemOpen [0]{}%
\providecommand \bibitemStop [0]{}%
\providecommand \bibitemNoStop [0]{.\EOS\space}%
\providecommand \EOS [0]{\spacefactor3000\relax}%
\providecommand \BibitemShut  [1]{\csname bibitem#1\endcsname}%
\let\auto@bib@innerbib\@empty
%</preamble>

\bibitem [{\citenamefont {Stewart}(2001)}]{Stewart01}%
  \BibitemOpen
  \bibfield  {author} {\bibinfo {author} {\bibfnamefont {G.~R.}\ \bibnamefont
  {Stewart}},\ }\href@noop {} {\bibfield  {journal} {\bibinfo  {journal}
  {Rev.~Mod.~Phys.}\ }\textbf {\bibinfo {volume} {73}},\ \bibinfo {pages} {797}
  (\bibinfo {year} {2001})}%%\BibitemShut {NoStop}%
  
\bibitem [{\citenamefont {v.~Lohneysen}\ \emph {et~al.}(2007)\citenamefont
  {v.~Lohneysen}, \citenamefont {Rosch}, \citenamefont {Vojta},\ and\
  \citenamefont {Wolfle}}]{Lohneysen07}%
  \BibitemOpen
  \bibfield  {author} {\bibinfo {author} {\bibfnamefont {H.}~\bibnamefont
  {v.~Lohneysen et al.}}, %\bibinfo {author} {\bibfnamefont {A.}~\bibnamefont {Rosch}},
 % \bibinfo {author} {\bibfnamefont {M.}~\bibnamefont {Vojta}}, \ and\ \bibinfo
 % {author} {\bibfnamefont {P.}~\bibnamefont {Wolfle}},
 \ }\href@noop {}
  {\bibfield  {journal} {\bibinfo  {journal} {Rev.~Mod.~Phys.}\ }\textbf
  {\bibinfo {volume} {79}},\ \bibinfo {pages} {1015} (\bibinfo {year}
  {2007})}%\BibitemShut {NoStop}%
  
\bibitem [{\citenamefont {Gegenwart}\ \emph {et~al.}(2008)\citenamefont
  {Gegenwart}, \citenamefont {Si},\ and\ \citenamefont
  {Steglich}}]{Gegenwart08}%
  \BibitemOpen
  \bibfield  {author} {\bibinfo {author} {\bibfnamefont {P.}~\bibnamefont
  {Gegenwart et al.}}, %\bibinfo {author} {\bibfnamefont {Q.}~\bibnamefont {Si}}, \
  %and\ \bibinfo {author} {\bibfnamefont {F.}~\bibnamefont {Steglich}},
  \ }\href@noop {} {\bibfield  {journal} {\bibinfo  {journal} {Nature~Physics}\
  }\textbf {\bibinfo {volume} {4}},\ \bibinfo {pages} {186} (\bibinfo {year}
  {2008})}%\BibitemShut {NoStop}%

  
\bibitem [{\citenamefont {Mathur}\ \emph {et~al.}(1998)\citenamefont {Mathur},
  \citenamefont {Grosche}, \citenamefont {Julian}, \citenamefont {Walker},
  \citenamefont {Freye}, \citenamefont {Haselwimmer},\ and\ \citenamefont
  {Lonzarich}}]{Grosche98}%
  \BibitemOpen
  \bibfield  {author} {\bibinfo {author} {\bibfnamefont {N.~D.}\ \bibnamefont
  {Mathur et al.}}, %\bibinfo {author} {\bibfnamefont {F.~M.}\ \bibnamefont {Grosche}},
  %\bibinfo {author} {\bibfnamefont {S.~R.}\ \bibnamefont {Julianet al.}}, \bibinfo
%  {author} {\bibfnamefont {I.~R.}\ \bibnamefont {Walkeret al.}}, \bibinfo {author}
%  {\bibfnamefont {D.~M.}\ \bibnamefont {Freyeet al.}}, \bibinfo {author}
%  {\bibfnamefont {R.~K.~W.}\ \bibnamefont {Haselwimmer}}, \ and\ \bibinfo
%  {author} {\bibfnamefont {G.~G.}\ \bibnamefont {Lonzarich}},
\ }\href@noop {}
  {\bibfield  {journal} {\bibinfo  {journal} {Nature}\ }\textbf {\bibinfo
  {volume} {394}},\ \bibinfo {pages} {39} (\bibinfo {year} {1998})}%\BibitemShut
  {NoStop}%
  
\bibitem [{\citenamefont {Monthoux}\ \emph {et~al.}(2007)\citenamefont
  {Monthoux}, \citenamefont {Pines},\ and\ \citenamefont
  {Lonzarich}}]{Monthoux07}%
  \BibitemOpen
  \bibfield  {author} {\bibinfo {author} {\bibfnamefont {P.}~\bibnamefont
  {Monthoux et al}}, 
  %\bibinfo {author} {\bibfnamefont {D.}~\bibnamefont {Pines}}, \
  %and\ \bibinfo {author} {\bibfnamefont {G.~G.}\ \bibnamefont {Lonzarich}},
  \ }\href@noop {} {\bibfield  {journal} {\bibinfo  {journal} {Nature}\ }\textbf
  {\bibinfo {volume} {450}},\ \bibinfo {pages} {1177} (\bibinfo {year}
  {2007})}%\BibitemShut {NoStop}%
  
\bibitem [{\citenamefont {Doniach}(1977)}]{Doniach77}%
  \BibitemOpen
  \bibfield  {author} {\bibinfo {author} {\bibfnamefont {S.}~\bibnamefont
  {Doniach}},\ }\href@noop {} {\bibfield  {journal} {\bibinfo  {journal}
  {Physica B \& C}\ }\textbf {\bibinfo {volume} {91}},\ \bibinfo {pages} {231}
  (\bibinfo {year} {1977})}%\BibitemShut {NoStop}%
  
\bibitem [{\citenamefont {Miyake}(2007)}]{Miyake07}%
  \BibitemOpen
  \bibfield  {author} {\bibinfo {author} {\bibfnamefont {K.}~\bibnamefont
  {Miyake}},\ }\href@noop {} {\bibfield  {journal} {\bibinfo  {journal}
  {J.~Phys.:~Condens.~Matter}\ }\textbf {\bibinfo {volume} {19}},\ \bibinfo
  {pages} {125201} (\bibinfo {year} {2007})}%\BibitemShut {NoStop}%
  
\bibitem [{\citenamefont {Watanabe}\ \emph {et~al.}(2008)\citenamefont
  {Watanabe}, \citenamefont {Tsuruta}, \citenamefont {Miyake},\ and\
  \citenamefont {Flouquet}}]{Watanabe08}%
  \BibitemOpen
  \bibfield  {author} {\bibinfo {author} {\bibfnamefont {S.}~\bibnamefont
  {Watanabe et al.}}, %\bibinfo {author} {\bibfnamefont {A.}~\bibnamefont {Tsuruta}},
  %\bibinfo {author} {\bibfnamefont {K.}~\bibnamefont {Miyake}}, \ and\ \bibinfo
 % {author} {\bibfnamefont {J.}~\bibnamefont {Flouquet}},
 \ }\href@noop {}
  {\bibfield  {journal} {\bibinfo  {journal} {Phys.~Rev.~Lett.}\ }\textbf
  {\bibinfo {volume} {100}},\ \bibinfo {pages} {236401} (\bibinfo {year}
  {2008})}%\BibitemShut {NoStop}%
  
\bibitem [{\citenamefont {Rueff}\ \emph {et~al.}(2011)\citenamefont {Rueff},
  \citenamefont {Raymond}, \citenamefont {Taguchi}, \citenamefont {Sikora},
  \citenamefont {Itie}, \citenamefont {Baudelet}, \citenamefont {Braithwaite},
  \citenamefont {Knebel},\ and\ \citenamefont {Jaccard}}]{Rueff11}%
  \BibitemOpen
  \bibfield  {author} {\bibinfo {author} {\bibfnamefont {J.-P.}\ \bibnamefont
  {Rueff et al.}}, %\bibinfo {author} {\bibfnamefont {S.}~\bibnamefont {Raymond}},
  %\bibinfo {author} {\bibfnamefont {M.}~\bibnamefont {Taguchiet al.}}, \bibinfo
 % {author} {\bibfnamefont {M.}~\bibnamefont {Sikoraet al.}}, \bibinfo {author}
%  {\bibfnamefont {J.-P.}\ \bibnamefont {Itieet al.}}, \bibinfo {author}
%  {\bibfnamefont {F.}~\bibnamefont {Baudeletet al.}}, \bibinfo {author}
%  {\bibfnamefont {D.}~\bibnamefont {Braithwaiteet al.}}, \bibinfo {author}
%  {\bibfnamefont {G.}~\bibnamefont {Knebel}}, \ and\ \bibinfo {author}
%  {\bibfnamefont {D.}~\bibnamefont {Jaccard}},
\ }\href@noop {} {\bibfield
  {journal} {\bibinfo  {journal} {Phys.~Rev.~Lett.}\ }\textbf {\bibinfo
  {volume} {106}},\ \bibinfo {pages} {186405} (\bibinfo {year}
  {2011})}%\BibitemShut {NoStop}%
  
   
\bibitem [{\citenamefont {Watanabe}\ and\ \citenamefont
  {Miyake}(2010)}]{Watanabe10}%
  \BibitemOpen
  \bibfield  {author} {\bibinfo {author} {\bibfnamefont {S.}~\bibnamefont
  {Watanabe}}\ and\ \bibinfo {author} {\bibfnamefont {K.}~\bibnamefont
  {Miyake}},\ }\href@noop {} {\bibfield  {journal} {\bibinfo  {journal}
  {Phys.~Rev.~Lett.}\ }\textbf {\bibinfo {volume} {105}},\ \bibinfo {pages}
  {186403} (\bibinfo {year} {2010})}%\BibitemShut {NoStop}%
  
  \bibitem [{\citenamefont {Ueda}\ and\ \citenamefont {Onuki}(1988)}]{heavy}%
  \BibitemOpen
  \bibfield  {author} {\bibinfo {author} {\bibfnamefont {K.}~\bibnamefont
  {Ueda}}\ and\ \bibinfo {author} {\bibfnamefont {Y.}~\bibnamefont {Onuki}},\
  }\href@noop {} {\emph {\bibinfo {title} {Physics of Heavy Electron
  Systems}}}\ (\bibinfo  {publisher} {Shokabo, Tokyo [in Japanese]},\ \bibinfo
  {year} {1988})%\BibitemShut {NoStop}%
  
\bibitem [{\citenamefont {Haen}\ \emph {et~al.}(1987)\citenamefont {Haen},
  \citenamefont {Flouquet}, \citenamefont {Lapierre}, \citenamefont {Lejay},\
  and\ \citenamefont {Remenyi}}]{Haen87}%
  \BibitemOpen
  \bibfield  {author} {\bibinfo {author} {\bibfnamefont {P.}~\bibnamefont
  {Haen et al.}}, 
  %\bibinfo {author} {\bibfnamefont {J.}~\bibnamefont {Flouquet}},
%  \bibinfo {author} {\bibfnamefont {F.}~\bibnamefont {Lapierreet al.}}, 
%\bibinfo  {author} {\bibfnamefont {P.}~\bibnamefont {Lejay}}, \ and\ \bibinfo {author}
%  {\bibfnamefont {G.}~\bibnamefont {Remenyi}},
\ }\href@noop {} {\bibfield
  {journal} {\bibinfo  {journal} {J.~Low~Temp.~Phys.}\ }\textbf {\bibinfo
  {volume} {67}},\ \bibinfo {pages} {391} (\bibinfo {year} {1987})}%\BibitemShut  {NoStop}%
  
\bibitem [{\citenamefont {Aoki}\ \emph {et~al.}(1993)\citenamefont {Aoki},
  \citenamefont {Uji}, \citenamefont {Albessard},\ and\ \citenamefont
  {Onuki}}]{Aoki93}%
  \BibitemOpen
  \bibfield  {author} {\bibinfo {author} {\bibfnamefont {H.}~\bibnamefont
  {Aoki et al.}}, %\bibinfo {author} {\bibfnamefont {S.}~\bibnamefont {Ujiet al.}}, \bibinfo
  %{author} {\bibfnamefont {A.~K.}\ \bibnamefont {Albessard}}, \ and\ \bibinfo
  %{author} {\bibfnamefont {Y.}~\bibnamefont {Onuki}},
  \ }\href@noop {}
  {\bibfield  {journal} {\bibinfo  {journal} {Phys.~Rev.~Lett.}\ }\textbf
  {\bibinfo {volume} {71}},\ \bibinfo {pages} {2110} (\bibinfo {year}
  {1993})}%\BibitemShut {NoStop}%
  
\bibitem [{\citenamefont {Watanabe}(2000)}]{Watanabe00}%
  \BibitemOpen
  \bibfield  {author} {\bibinfo {author} {\bibfnamefont {K.}~\bibnamefont
  {Watanabe}},\ }\href@noop {} {\bibfield  {journal} {\bibinfo  {journal}
  {J.~Phys.~Soc.~Jpn.}\ }\textbf {\bibinfo {volume} {69}},\ \bibinfo {pages}
  {2947} (\bibinfo {year} {2000})}%\BibitemShut {NoStop}%
  
\bibitem [{\citenamefont {Miyake}\ and\ \citenamefont
  {Ikeda}(2006)}]{MiyakeIkeda06}%
  \BibitemOpen
  \bibfield  {author} {\bibinfo {author} {\bibfnamefont {K.}~\bibnamefont
  {Miyake}}\ and\ \bibinfo {author} {\bibfnamefont {H.}~\bibnamefont {Ikeda}},\
  }\href@noop {} {\bibfield  {journal} {\bibinfo  {journal}
  {J.~Phys.~Soc.~Jpn.}\ }\textbf {\bibinfo {volume} {75}},\ \bibinfo {pages}
  {033704} (\bibinfo {year} {2006})}%\BibitemShut {NoStop}%
  
\bibitem [{\citenamefont {Daou}\ \emph {et~al.}(2006)\citenamefont {Daou},
  \citenamefont {Bergemann},\ and\ \citenamefont {Julian}}]{Daou06}%
  \BibitemOpen
  \bibfield  {author} {\bibinfo {author} {\bibfnamefont {R.}~\bibnamefont
  {Daou et al.}}, 
  %\bibinfo {author} {\bibfnamefont {C.}~\bibnamefont {Bergemann}}, \
  %and\ \bibinfo {author} {\bibfnamefont {S.~R.}\ \bibnamefont {Julian}},
  \ }\href@noop {} {\bibfield  {journal} {\bibinfo  {journal} {Phys.~Rev.~Lett.}\
  }\textbf {\bibinfo {volume} {96}},\ \bibinfo {pages} {026401} (\bibinfo
  {year} {2006})}%\BibitemShut {NoStop}%
  
\bibitem [{\citenamefont {Sugi}\ \emph {et~al.}(2008)\citenamefont {Sugi},
  \citenamefont {Matsumoto}, \citenamefont {Kimura}, \citenamefont
  {Komatsubara}, \citenamefont {Aoki}, \citenamefont {Terashima},\ and\
  \citenamefont {Uji}}]{Sugi08}%
  \BibitemOpen
  \bibfield  {author} {\bibinfo {author} {\bibfnamefont {M.}~\bibnamefont
  {Sugi et al.}}, %\bibinfo {author} {\bibfnamefont {Y.}~\bibnamefont {Matsumoto}},
  %\bibinfo {author} {\bibfnamefont {N.}~\bibnamefont {Kimuraet al.}}, \bibinfo
 % {author} {\bibfnamefont {T.}~\bibnamefont {Komatsubaraet al.}}, \bibinfo {author}
 % {\bibfnamefont {H.}~\bibnamefont {Aokiet al.}}, \bibinfo {author} {\bibfnamefont
%  {T.}~\bibnamefont {Terashima}}, \ and\ \bibinfo {author} {\bibfnamefont
%  {S.}~\bibnamefont {Uji}},
\ }\href@noop {} {\bibfield  {journal} {\bibinfo
  {journal} {Phys.~Rev.~Lett.}\ }\textbf {\bibinfo {volume} {101}},\ \bibinfo
  {pages} {056401} (\bibinfo {year} {2008})}%\BibitemShut {NoStop}%
  
\bibitem [{\citenamefont {Matsuda}\ \emph {et~al.}(2007)\citenamefont
  {Matsuda}, \citenamefont {Inami}, \citenamefont {Ohwada}, \citenamefont
  {Murata}, \citenamefont {Nojiri}, \citenamefont {Murakami}, \citenamefont
  {Ohta}, \citenamefont {Zhang},\ and\ \citenamefont {Yoshimura}}]{Matsuda07}%
  \BibitemOpen
  \bibfield  {author} {\bibinfo {author} {\bibfnamefont {Y.~H.}\ \bibnamefont
  {Matsuda et al.}}, 
  %\bibinfo {author} {\bibfnamefont {T.}~\bibnamefont {Inami}},
  %\bibinfo {author} {\bibfnamefont {K.}~\bibnamefont {Ohwadaet al.}}, \bibinfo
 % {author} {\bibfnamefont {Y.}~\bibnamefont {Murataet al.}}, \bibinfo {author}
%  {\bibfnamefont {H.}~\bibnamefont {Nojiriet al.}}, \bibinfo {author} {\bibfnamefont
%  {Y.}~\bibnamefont {Murakamiet al.}}, \bibinfo {author} {\bibfnamefont
%  {H.}~\bibnamefont {Ohtaet al.}}, \bibinfo {author} {\bibfnamefont {W.}~\bibnamefont
%  {Zhang}}, \ and\ \bibinfo {author} {\bibfnamefont {K.}~\bibnamefont
%  {Yoshimura}},
\ }\href@noop {} {\bibfield  {journal} {\bibinfo  {journal} {J.\
  Phys.\ Soc.\ Jpn.}\ }\textbf {\bibinfo {volume} {76}},\ \bibinfo {pages}
  {034702} (\bibinfo {year} {2007})}%\BibitemShut {NoStop}%
  
\bibitem [{\citenamefont {Rueff}\ \emph {et~al.}(2006)\citenamefont {Rueff},
  \citenamefont {Itie}, \citenamefont {Taguchi}, \citenamefont {Hague},
  \citenamefont {Mariot}, \citenamefont {Delaunay}, \citenamefont {Kappler},\
  and\ \citenamefont {Jaouen}}]{Rueff06}%
  \BibitemOpen
  \bibfield  {author} {\bibinfo {author} {\bibfnamefont {J.~P.}\ \bibnamefont
  {Rueff et al.}}, %\bibinfo {author} {\bibfnamefont {J.~P.}\ \bibnamefont {Itie}},
 % \bibinfo {author} {\bibfnamefont {M.}~\bibnamefont {Taguchiet al.}}, \bibinfo
%  {author} {\bibfnamefont {C.~F.}\ \bibnamefont {Hagueet al.}}, \bibinfo {author}
 % {\bibfnamefont {J.~M.}\ \bibnamefont {Mariotet al.}}, \bibinfo {author}
 % {\bibfnamefont {R.}~\bibnamefont {Delaunayet al.}}, \bibinfo {author}
%  {\bibfnamefont {J.~P.}\ \bibnamefont {Kappler}}, \ and\ \bibinfo {author}
%  {\bibfnamefont {N.}~\bibnamefont {Jaouen}},
\ }\href@noop {} {\bibfield
  {journal} {\bibinfo  {journal} {Phys.~Rev.~Lett.}\ }\textbf {\bibinfo
  {volume} {96}},\ \bibinfo {pages} {237403} (\bibinfo {year}
  {2006})}%\BibitemShut {NoStop}%
  
\bibitem [{\citenamefont {Yano}\ \emph {et~al.}(2008)}]{yano08}%
\BibitemOpen
\bibfield  {author} {\bibinfo {author} {\bibfnamefont {M.}\ \bibnamefont
{Yano et al.}}, 
\ }\href@noop {} {\bibfield
{journal} {\bibinfo  {journal} {Phys.~Rev.~B}\ }\textbf {\bibinfo
{volume} {77}},\ \bibinfo {pages} {035118} (\bibinfo {year}
{2008})}


\bibitem [{\citenamefont {Willers}\ \emph {et~al.}(2012)}]{willers12}%
\BibitemOpen
\bibfield  {author} {\bibinfo {author} {\bibfnamefont {T.}\ \bibnamefont
{Willers et al.}}, 
\ }\href@noop {} {\bibfield
{journal} {\bibinfo  {journal} {Phys.~Rev.~B}\ }\textbf {\bibinfo
{volume} {85}},\ \bibinfo {pages} {035117} (\bibinfo {year}
{2012})}

  
  
\bibitem [{\citenamefont {Watanabe}\ \emph {et~al.}(2009)\citenamefont
  {Watanabe}, \citenamefont {Tsuruta}, \citenamefont {Miyake},\ and\
  \citenamefont {Flouquet}}]{Watanabe09}%
  \BibitemOpen
  \bibfield  {author} {\bibinfo {author} {\bibfnamefont {S.}~\bibnamefont
  {Watanabe et al.}}, %\bibinfo {author} {\bibfnamefont {A.}~\bibnamefont {Tsuruta}},
 % \bibinfo {author} {\bibfnamefont {K.}~\bibnamefont {Miyake}}, \ and\ \bibinfo
 % {author} {\bibfnamefont {J.}~\bibnamefont {Flouquet}},
 \ }\href@noop {}
  {\bibfield  {journal} {\bibinfo  {journal} {J.\ Phys.\ Soc.\ Jpn.}\ }\textbf
  {\bibinfo {volume} {78}},\ \bibinfo {pages} {104706} (\bibinfo {year}
  {2009})}%\BibitemShut {NoStop}%
  
\bibitem [{\citenamefont {Paulsen}\ \emph {et~al.}(1990)\citenamefont
  {Paulsen}, \citenamefont {Lacerda}, \citenamefont {Puech}, \citenamefont
  {Haen}, \citenamefont {Lejay}, \citenamefont {Tholence}, \citenamefont
  {Flouquet},\ and\ \citenamefont {de~Visser}}]{Paulsen90}%
  \BibitemOpen
  \bibfield  {author} {\bibinfo {author} {\bibfnamefont {C.}~\bibnamefont
  {Paulsen et al.}}, 
  %\bibinfo {author} {\bibfnamefont {A.}~\bibnamefont {Lacerda}},
 % \bibinfo {author} {\bibfnamefont {L.}~\bibnamefont {Puechet al.}}, \bibinfo
%  {author} {\bibfnamefont {P.}~\bibnamefont {Haenet al.}}, \bibinfo {author}
  %{\bibfnamefont {P.}~\bibnamefont {Lejayet al.}}, \bibinfo {author} {\bibfnamefont
 % {J.~L.}\ \bibnamefont {Tholenceet al.}}, \bibinfo {author} {\bibfnamefont
 % {J.}~\bibnamefont {Flouquet}}, \ and\ \bibinfo {author} {\bibfnamefont
%  {A.}~\bibnamefont {de~Visser}},
\ }\href@noop {} {\bibfield  {journal}
  {\bibinfo  {journal} {J.~Low.~Phys.}\ }\textbf {\bibinfo {volume} {81}},\
  \bibinfo {pages} {317} (\bibinfo {year} {1990})}%\BibitemShut {NoStop}%
  
\bibitem [{\citenamefont {Grier}\ \emph {et~al.}(1981)\citenamefont {Grier},
  \citenamefont {Parks}, \citenamefont {Shapiro},\ and\ \citenamefont
  {Majkrzak}}]{Grier81}%
  \BibitemOpen
  \bibfield  {author} {\bibinfo {author} {\bibfnamefont {B.~H.}\ \bibnamefont
  {Grier et al.}}, %\bibinfo {author} {\bibfnamefont {R.~D.}\ \bibnamefont {Parks}},
%  \bibinfo {author} {\bibfnamefont {S.~M.}\ \bibnamefont {Shapiro}}, \ and\
%  \bibinfo {author} {\bibfnamefont {C.~F.}\ \bibnamefont {Majkrzak}},
\ }\href@noop {} {\bibfield  {journal} {\bibinfo  {journal} {Phys.~Rev.~B}\
  }\textbf {\bibinfo {volume} {24}},\ \bibinfo {pages} {6242} (\bibinfo {year}
  {1981})}%\BibitemShut {NoStop}%
  
\bibitem [{\citenamefont {Matsuda}\ \emph {et~al.}(2012)\citenamefont
  {Matsuda}, \citenamefont {Nakamura}, \citenamefont {Her}, \citenamefont
  {Kindo}, \citenamefont {Michimura}, \citenamefont {Inami}, \citenamefont
  {Mizumaki}, \citenamefont {Kawamura}, \citenamefont {Suzuki}, \citenamefont
  {Chen}, \citenamefont {Ohta},\ and\ \citenamefont {Yoshimura}}]{Matsuda12}%
  \BibitemOpen
  \bibfield  {author} {\bibinfo {author} {\bibfnamefont {Y.~H.}\ \bibnamefont
  {Matsuda et al.}}, 
  %\bibinfo {author} {\bibfnamefont {T.}~\bibnamefont {Nakamura}},
  %\bibinfo {author} {\bibfnamefont {J.~L.}\ \bibnamefont {Heret al.}}, \bibinfo
 % {author} {\bibfnamefont {K.}~\bibnamefont {Kindoet al.}}, \bibinfo {author}
%  {\bibfnamefont {S.}~\bibnamefont {Michimuraet al.}}, \bibinfo {author}
%  {\bibfnamefont {T.}~\bibnamefont {Inamiet al.}}, \bibinfo {author} {\bibfnamefont
%  {M.}~\bibnamefont {Mizumakiet al.}}, \bibinfo {author} {\bibfnamefont
%  {N.}~\bibnamefont {Kawamuraet al.}}, \bibinfo {author} {\bibfnamefont
%  {M.}~\bibnamefont {Suzukiet al.}}, \bibinfo {author} {\bibfnamefont
%  {B.}~\bibnamefont {Chenet al.}}, \bibinfo {author} {\bibfnamefont {H.}~\bibnamefont
%  {Ohta}}, \ and\ \bibinfo {author} {\bibfnamefont {K.}~\bibnamefont
%  {Yoshimura}},
\ }\href@noop {} {\bibfield  {journal} {\bibinfo  {journal} {J.\
  Phys.\ Soc.\ Jpn.}\ }\textbf {\bibinfo {volume} {81}},\ \bibinfo {pages}
  {015002} (\bibinfo {year} {2012})}%\BibitemShut {NoStop}%
  
\bibitem [{\citenamefont {Sakakibara}\ \emph {et~al.}(1995)\citenamefont
  {Sakakibara}, \citenamefont {Tayama}, \citenamefont {Matsuhira},
  \citenamefont {Mitamura}, \citenamefont {Amitsuka}, \citenamefont {Maezawa},\
  and\ \citenamefont {Onuki}}]{Sakakibara95}%
  \BibitemOpen
  \bibfield  {author} {\bibinfo {author} {\bibfnamefont {T.}~\bibnamefont
  {Sakakibara et al.}}, %\bibinfo {author} {\bibfnamefont {T.}~\bibnamefont {Tayama}},
 % \bibinfo {author} {\bibfnamefont {K.}~\bibnamefont {Matsuhiraet al.}}, \bibinfo
%  {author} {\bibfnamefont {H.}~\bibnamefont {Mitamuraet al.}}, \bibinfo {author}
%  {\bibfnamefont {H.}~\bibnamefont {Amitsukaet al.}}, \bibinfo {author}
%  {\bibfnamefont {K.}~\bibnamefont {Maezawa}}, \ and\ \bibinfo {author}
%  {\bibfnamefont {Y.}~\bibnamefont {Onuki}},
\ }\href@noop {} {\bibfield
  {journal} {\bibinfo  {journal} {Phys.~Rev.~B}\ }\textbf {\bibinfo {volume}
  {51}},\ \bibinfo {pages} {12030} (\bibinfo {year} {1995})}
  %\BibitemShut{NoStop}%
  
\end{thebibliography}

%

\end{document}